# A THEORETICAL STUDY OF ELASTIC, ELECTRONIC, OPTICAL AND THERMODYNAMIC PROPERTIES OF $AlB_2$ AND $TaB_2$

N. JAHAN[*], M. A. ALI

*Department of Physics, Chittagong University of Engineering and Technology, Chittagong, Bangladesh.*

## ABSTRACT

Using ab-initio method we have studied the structural, elastic, electronic, optical and thermodynamic properties of $AlB_2$ and $TaB_2$. We have used plane wave pseudopotential with generalised gradient approximation as implemented in CASTEP program. The independent elastic constants, bulk modulus, Young's modulus, shear modulus, anisotropic factor, Pugh ratio, etc are calculated. This results show that $AlB_2$ and $TaB_2$ are mechanically stable. We have also calculated the band structure and density of states. Band structure results show that $AlB_2$ and $TaB_2$ show metallic behaviour. In order to understand the electronic properties in a better way we have also calculated dielectric function, refractive index, absorption coefficient, conductivity, loss function and reflectivity. Results of absorption coefficient and conductivity are in good agreement with the band structure results. The effect of temperature and pressure on the bulk modulus, Debye temperature, specific heat and also on the thermal expansion coefficient are derived from the quasi-harmonic Debye model with phononic effect in both cases. The results for the elastic and electronic properties are compared to experimental measurements and with the results obtained in different band-structure calculations, however, no optical data are available for comparison.

**Keywords:** First-principles; Elastic properties; Electronic band structure; Optical properties.

## 1. INTRODUCTION

Transition metal diborides have been received a great deal of interest for many years, due to their unique physical and chemical properties such as hardness, high melting point, chemical inertness, etc., and belong to the most potential materials for industrial applications [1-12]. Moreover, the discovery of superconductivity in $MgB_2$ at $T_C \sim 39K$ [13] has led to an extensive search in finding superconductivity in other diborides with hexagonal $AlB_2$-type structure. One of the aluminium-rich phases of the binary aluminium-boron system is $AlB_2$. Among the intermetallic phases, $AlB_2$ is one of the most frequently observed materials for a long time. The crystal structure is simple which contains graphite like boron layers separated by aluminium atoms in hexagonal prismatic voids (space group *P6/mmm*). For the chemical variability and simplicity of the crystal structure, $AlB_2$ is being a very interesting one for the systematic investigation of crystal structure, chemical and physical properties to the researcher. $TaB_2$ is a transitional metal diboride which belongs to the hexagonal system with $AlB_2$-like structure (P6/mmm) [14].

---

[*]Corresponding author: Email: nusrat.jahan83@yahoo.com

Previous investigations of $AlB_2$ and $TaB_2$, theoretically and experimentally, have been reported in literature [15-20]. Specifically, the structural and elastic properties for $AlB_2$ and $TaB_2$ have been addressed by Shein and Ivanovskii [16]. Zhou Xiao-Lin studied the structural and thermodynamic properties of $AlB_2$ compound [14].Recently, the structural, elastic and electronic properties of $TaB_2$ with $AlB_2$ structure under high pressure have been studied by Zhang *et al* [18]. Moreover, the structural, mechanical and electronic properties of $TaB_2$ have also been studied by Zhao *et al*. [19]. Furthermore, Vajeeston *et al* studied the electronic structure, bonding, and ground-state properties of $AlB_2$-type transition-metal diborides [20]. However, most of them dealt with the structural, elastic and electronic properties of $AlB_2$ and $TaB_2$ but the optical properties which are very important to understand the electronic properties in better way were not investigated yet. In this paper, we have tried to explain the calculated results of structural, elastic, electronic, optical and thermodynamic properties of $AlB_2$ and $TaB_2$ and then compared our results with previous results where available.

## 2. METHODOLOGY

The first-principles calculations were performed by employing pseudo-potential plane-waves (PP-PW) approach based on the density functional theory (DFT) [21, 22] and implemented in the CASTEP code [23]. The major advantages of this approach are: the ease of computing forces and stresses; good convergence control with respect to all employed computational parameters; favorable scaling with number of atoms in the system and the ability to make cheaper calculations by neglecting core electrons. The exchange-correlation potential is treated within the GGA due to Perdew, Burke and Ernzerhoff (GGA-PBE) [24]. The presence of tightly-bound core electrons was represented by non local ultra-soft pseudopotentials of the Vanderbilt-type [25]. The states, Al $3s^2, 3p^1$, Ta $6s^2, 5d^3$ and B $2s^2 2p^1$ were treated as valence states. The two parameters that affect the accuracy of calculations are the kinetic energy cut-off which determines the number of plane waves in the expansion and the number of special k-points used for the Brillouin zone (BZ) integration. We performed convergence with respect to BZ sampling and the size of the basis set. Converged results were achieved with $15 \times 15 \times 15$ and $10 \times 10 \times 9$ special k-points mesh [26] for $AlB_2$ and $TaB_2$, respectively. The size of the basis set is given by cut-off energy equal to 550 eV for both cases. The structural parameters were determined using the BFGS [27] minimization technique.

## 3. RESULTS AND DISCUSSION

### *3.1 Structural and elastic properties*:

$AlB_2$ and $TaB_2$ are belonging to the hexagonal crystal system. The space group is P6/mmm (191). The equilibrium structure of the $AlB_2$ and $TaB_2$ are obtained by minimizing its unit cell with respect to the total energy. The position of atoms in the unit cell are: Al/Ta atom is at (0, 0, 0) and the 2B atoms are at (1/3, 2/3, 1/2) and (2/3, 1/3, 1/2). Each boron atom here is equidistance from three other boron atoms. The optimized lattice parameters are shown in Table 1. Our results are in good agreement with the theoretical and experimental results [16, 28]. The calculated independent elastic constants are also included in Table 1. For hexagonal crystals, the mechanical stability requires the elastic constants satisfying the well-known Born stability criteria [29]: $C_{11}>0$; $C_{11}-C_{12}>0$; $C_{44}>0$, $(C_{11} + C_{12}) C_{33} - 2C^2_{13}>0$. From our calculated $C_{ij}$ shown in Table 1, it is known that the hexagonal $AlB_2$ and $TaB_2$ are mechanically stable.

The calculated elastic parameters (bulk modulus $B$, compressibility $K$, shear modulus $G$, Young's modulus $E$, and Poisson's ratio $v$) of $AlB_2$ and $TaB_2$ are also given in Table 1. The expression for Y and $v$ can be found else-where [30]. The arithmetic average of the Voigt ($B_V$, $G_V$) and the Reuss ($B_R$, $G_R$) bounds is used to estimate the polycrystalline modulus. In the terms of the Voigt–Reuss–Hill approximations [31]: $B_H \equiv B = \frac{1}{2}(B_R+B_V)$ and $G_H \equiv G = 1/2(G_R+G_V)$, where $B$ and $G$ represent the bulk modulus and shear modulus respectively. In addition, Zener's anisotropy index $A = 2C_{44}/(C_{11} - C_{12})$ [32] and the so called Pugh ratio $B/G$ [33] are also calculated. Taking an overall look at the phases featured in Table 1. The bulk modulus ($B$) of a substance measures the substance's resistance to uniform compression. From table 1 we see that $TaB_2$ has a greater bulk modulus than $AlB_2$ which indicates that more pressure is required to change in volume for $TaB_2$ than $AlB_2$. The inverse of bulk modulus is the compressibility, which are also given in Table 1. We can see from table that $B > G$ for both cases, therefore the limiting parameter for stability of these compounds is the shear modulus $G$.

**Table 1:** The optimized structural parameters, independent elastic constants $C_{ij}$, bulk modulus $B$, compressibility $K$, shear modulus $G$, Young's modulus $Y$, Poisson ratio $v$, Zenger's anisotropy index $A$, and Pugh ratio $B/G$ of $AlB_2$ and $TaB_2$

| Parameters | $AlB_2$ | $TaB_2$ |
|---|---|---|
| $a$ (Å) | 3.006, 3.005[a], | 3.137, 3.088[a], |
| $c$ (Å) | 3.254, 3.257[a], | 3.344, 3.241[a], |
| $C_{11}$ (GPa) | 524, 530[b], 665[c] | 596, 708[b], 712[d] |
| $C_{12}$ (GPa) | 107, 82[b], 41[c] | 141, 129[b], 130[d] |
| $C_{13}$ (GPa) | 24, 67[b], 17[c] | 196, 218[b], 217[d] |
| $C_{33}$ (GPa) | 333, 272[b], 417[c] | 431, 517[b], 525[d] |
| $C_{44}$ (GPa) | 23, 32[b], 58[c] | 191, 236[b], 238[d] |
| $B$ (GPa) | 180, 196[b] | 297, 340[b], 341[d] |
| $K$ (GPa$^{-1}$) | 0.00575 | 0.00338 |
| $G$ (GPa) | 132, 132[b] | 191, 243[b], 241[d] |
| $Y$ (GPa) | 231, 250[b] | 485, 600[b], 586[d] |
| $v$ | 0.20, 0.27[b] | 0.23, 0.23[b], 0.21[d] |
| $A$ | 0.11, 0.51[b] | 0.83, 0.73[b] |
| $B/G$ | 1.36, 1.49[b] | 1.55, 1.36[b], 1.41[d] |

[a]ref. 28, [b]ref. 16, [c]ref. 17, [d]ref. 18

The Young's modulus $Y$ measures the response to a uniaxial stress averaged over all directions and is used often to denote a measure of stiffness, *i.e.* the larger is the value of $Y$, the stiffer is the material. The large values of $Y$ for both cases indicate that they will be stiff [34]. An additional argument for the variation in the brittle/ductile behavior of the examined phases follows from the calculated Poisson's ratio $v$, Table 1. Indeed, for brittle materials these values are small enough ($v \sim 0.1$), whereas for ductile metallic materials $v$ is typically 0.33. We can see that the examined compound lies in between the brittle/ductile border line. Let us consider elastic anisotropy parameter. For this purpose we have estimated Zener's anisotropy index $A$. For isotropic case $A = 1$, while the deviations from unity measure the degree of elastic anisotropy. Our results show that the compound under consideration is highly anisotropic. One of the most widely used malleability indicators of materials is the Pugh's ductility index ($B/G$ ratio) . As is known, if $B/G < 1.75$ the material will behave in a brittle manner and if $B/G > 1.75$, the material demonstrates ductileness. The compounds under considerstion will behave in a brittle manner.

*3.2 Electronic properties*:

The calculated energy band structure for $TaB_2$ and $AlB_2$, at equilibrium lattice parameters, along the high symmetry directions in the Brillouin zone are shown in Fig. 1(a) and (b). The Fermi level is chosen to be zero of the energy scale. The valence and conduction bands overlap considerably and there is no band gap at the Fermi level. As a result $TaB_2$ and $AlB_2$ will exhibit metallic properties.

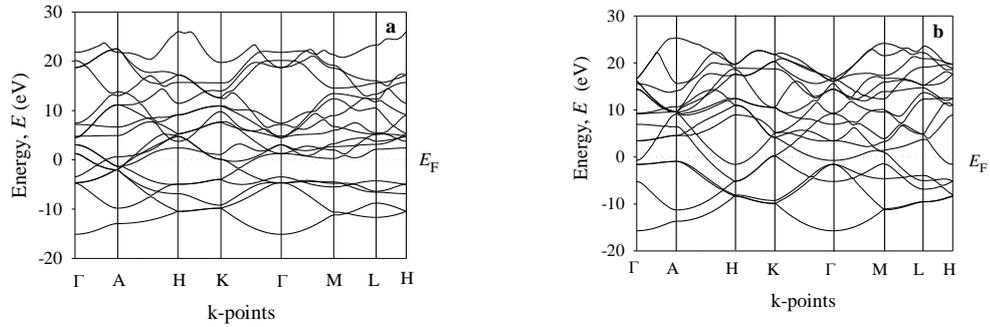

**Fig-1:** Band structure of  (a) $TaB_2$ and (b) $AlB_2$.

Moreover, in order to elucidate the different contributions from the different components in the materials to the conductivity we also calculate total and partial density of states as shown in Fig.2 (a) and (b). The DOS values of $AlB_2$ and $TaB_2$ are 0.43 and 0.5 at the Fermi level which are in good agreement with Vajeeston *et al* [20] and Zhang *et al* [18], respectively. For $AlB_2$, in Fig. 2(b) we see that the contribution from Al 3*s*, Al 3*p* and B 2*p* are almost equal to the Fermi level i.e. to the conduction properties. On the other hand for $TaB_2$, in Fig. 2(a), we observe that Ta 5*d* electrons are mainly contributing to the DOS at the Fermi level, and should be involved in the conduction properties. The contribution from B 2*p* states is noticeable but an order of magnitude smaller than that of Ta 5*d* states. For both cases B 2*s* do not contribute at the Fermi level and therefore is not involved in the conduction properties. The energy states below the Fermi level are due to the hybridizing of Al 3*s*, Al 2*p*, B 2*s* and B 2*p* states for $AlB_2$ and Ta 6*s*, Ta 5*d*, B2*s* and B 2*p* states for $TaB_2$ with a dominant contribution from B 2*p* states.

*3.3 Optical Properties*

The optical properties of $AlB_2$ and $TaB_2$ may be obtained from the complex dielectric function, $\varepsilon(\omega) = \varepsilon_1(\omega) + i\varepsilon_2(\omega)$. The imaginary part $\varepsilon_2(\omega)$ is obtained from the momentum matrix elements between the occupied and the unoccupied electronic states and calculated directly by CASTEP [35] using the following equation:

$$\varepsilon_2(\omega) = \frac{2e^2\pi}{\Omega\varepsilon_0} \sum_{k,v,c} \left|\psi_k^c|\hat{u}.r|\psi_k^v\right|^2 \delta(E_k^c - E_k^v - E)$$

where $\hat{u}$ is the vector defining the polarization of the incident electric field. ω is the light frequency, e is the electronic charge and $\psi_k^c$ and $\psi_k^v$ are the conduction and valence band wave functions at k, respectively. The real part is derived from the imaginary part $\varepsilon_2(\omega)$ by the Kramers–Kronig transform. All other optical constants such as refractive index, absorption spectrum, loss-function, reflectivity and conductivity are those given by Eqs. 49–54 in Ref. [35].

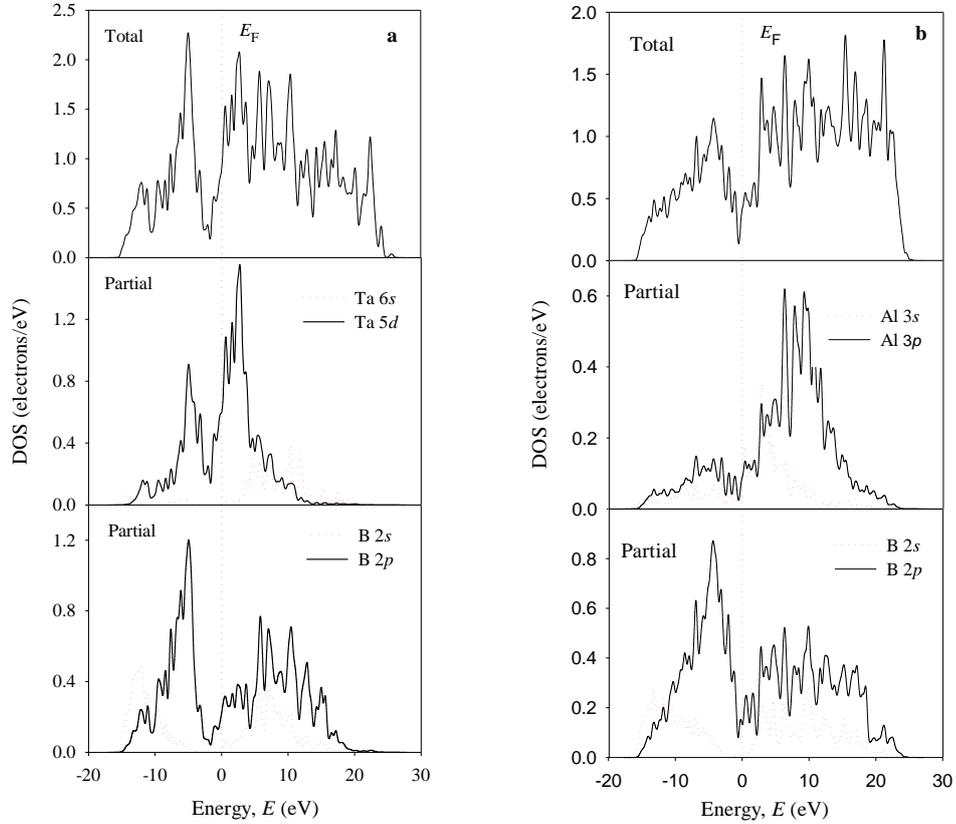

**Fig-2:** Total and Partial Density of States of (a) $TaB_2$ and (b) $AlB_2$.

The optical properties of solids provide an important tool for studying electronic properties of materials. Fig. 3 and 4 (a)–(f) shows the calculated optical properties of $TaB_2$ and $AlB_2$ from the polarization vector (1 0 0) for photon energies up to 40 eV. In our calculation, we used a Gaussian smearing which is 0.5 eV. This smears out the fermi level, so that more k-points will be effectively on the fermi surface. The complex dielectric function is intimately connected to band structure for solid. The primary quantity which is the probability of photon absorption for any crystalline material is directly related to the imaginary part of the optical dielectric function ε (ω).

The imaginary part $\varepsilon_2(\omega)$ of the dielectric function has the peaks at 2.9 and 6.3 eV for $AlB_2$ [Fig. 4 (a)] are due to transitions of electrons from B $2p$ VB to Al $3p$ CB but there is only one peak at 0.46 eV for $TaB_2$ due to transitions of electrons from B $2p$ VB to Ta $5d$ CB

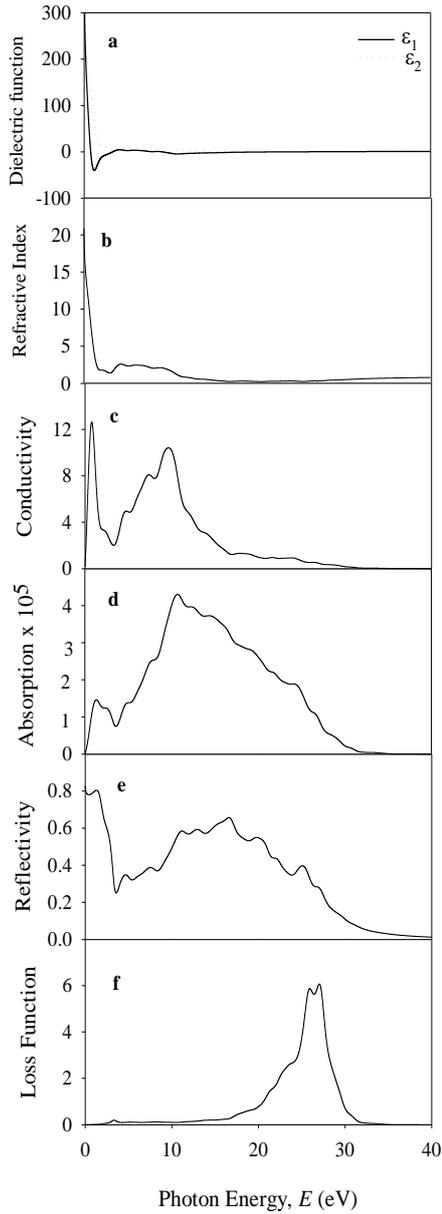

**Fig-3:** Optical properties of $TaB_2$.

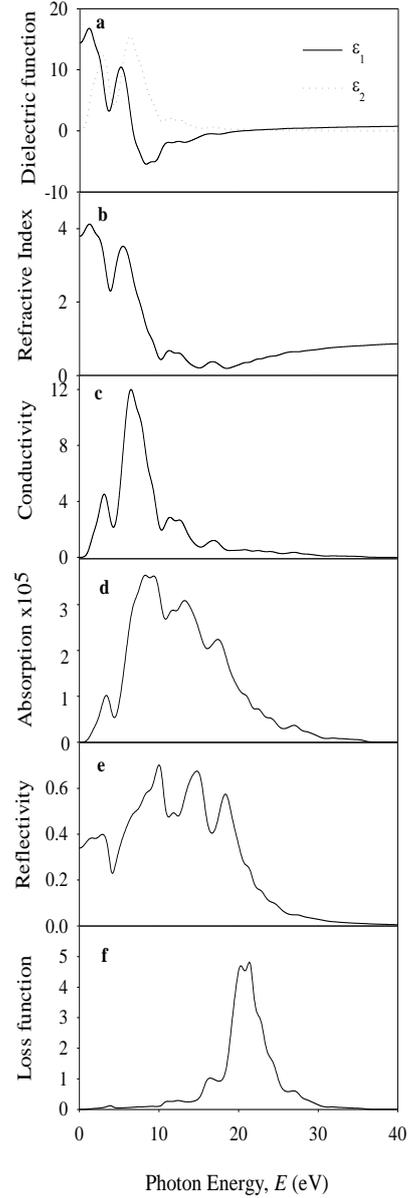

**Fig-4:** Optical properties of $AlB_2$.

The refractive indexes of the compounds under consideration are displayed in Figs. 3 and 4 (b). The static values are of 19.7 and 3.7 for $TaB_2$ and $AlB_2$, respectively. In case of $AlB_2$, there are two peaks at 1.3 and 5.5 eV. On the other hand for $TaB_2$, there is an only one peak at ~4.12 eV. Photoconductivity is the increase in electrical conductivity that results from the increase in the number of free carriers due to the absorption of photons of sufficient frequency. We can see from the Figs. 3 and 4 (c), the photoconductivity starts with a non-zero value corresponds to the incident photon energy (for $AlB_2$, at 0.01 eV this has a very small value ~ 0.731, so it is hardly seen) which shows that the compounds have no band gap. Two main peaks are observed at 0.76 and 9.6 eV for $TaB_2$ and at 3.1 and 6.4 eV for $AlB_2$. Figs. 3 and 4 (d) show non-zero absorption coefficient due to the metallic nature of $TaB_2$ and $AlB_2$. The corresponding peaks of conductivity are also seen in absorption spectra. For $TaB_2$, this spectrum also rises below 11 eV with two peaks at 1.3 and 10.6 eV and then decreased rapidly to 0 at 34 eV. For $AlB_2$, this spectrum rises sharply below 7 eV with two peaks at 3.3 and 6.4 eV and then decreased rapidly to 0 at ~35 eV. There are also two small peaks at 13.3 and 17.6 eV energy.

Figs. 3 and 4 (e) illustrates the dependence of reflectivity on frequency, exhibiting the dramatic discontinuous drop in $R$ at $\omega = \omega p$, which has come to be known as the plasma reflection edge. For $AlB_2$ [Fig. 4 (e)] we found that it can reflects only 37% of incident light at low frequency but it can less likely be a good reflector in the range of 10-20 eV photon energy. On the other hand $TaB_2$ can reflects almost 80% of incident light in low energy region that indicates its applicability as a candidate material for use to remove solar heating. The function $L(\omega)$ shown in Figs. 3 and 4 (f), describes the energy loss of a fast electron traversing in the material. Its peak is defined as the bulk plasma frequency $\omega_\rho$, which occurs at $\varepsilon_2 \ll 1$ and $\varepsilon_1 = 0$ [36, 37]. Moreover, the positions of peaks in $L(\omega)$ spectra, which correspond to the so-called plasma frequency, point out the transition from the metallic property [$\varepsilon_1(\omega) < 0$] to the dielectric property [$\varepsilon_1(\omega) > 0$] for a material. In the energy-loss spectrum, we see that the plasma frequency $\omega_\rho$ of $TaB_2$ is 26 eV and for $AlB_2$ it is equal to ~21.2 eV. When the incident light's frequency is higher than that of plasma frequency, the material becomes transparent. In addition, the peaks of $L(\omega)$ correspond to the trailing edges in the reflection spectra as shown in Fig. 3 and 4 (e and f).

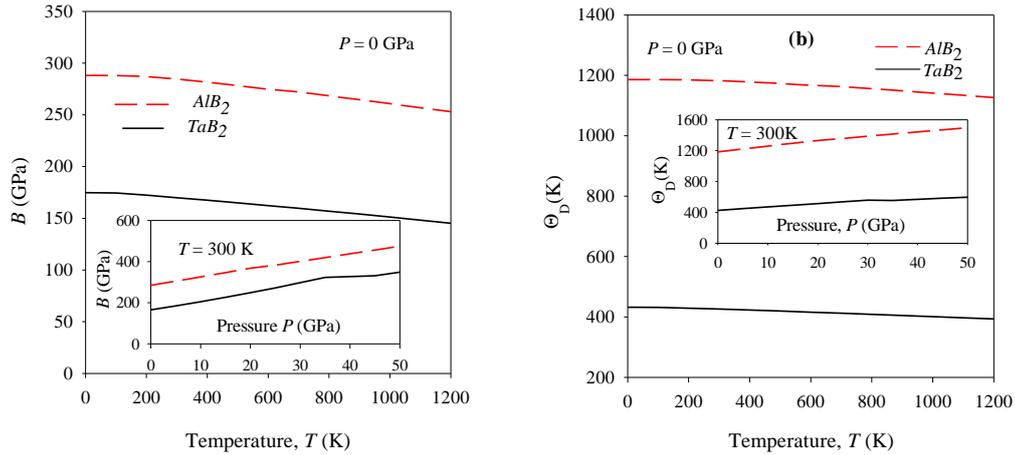

**Fig-5:** Temperature dependence of (a) Bulk modulus and (b) Debye temperature of $TaB_2$ and $AlB_2$. *Inset* shows pressure variation.

*3.4 Thermodynamical properties*

We investigate the thermodynamics properties of $TaB_2$ and $AlB_2$ by using the quasi-harmonic Debye model in a manner as described elsewhere [38, 39]. Here we computed the bulk modulus, Debye temperature, specific heats and volume thermal expansion coefficient at different temperatures and pressures. For this we utilized E-V data obtained from the third order Birch-Murnaghan equation of state [40] using zero temperature and zero pressure equilibrium values of $E_0$, $V_0$, $B_0$, based on DFT method. Fig.5 (a) represents the temperature variation of isothermal bulk modulus of $TaB_2$ and $AlB_2$, the inset of which shows bulk modulus *B* data as a function of pressure. Our results reveal that all *B* values are nearly flat below 100K. Above 100K, B for $AlB_2$ decreases at a slightly faster rate than the other compound. The inset shows the pressure variation of bulk modulus at 300 K. Further the bulk modulus, signifying the average strength of the coupling between the neighboring atoms, increases with pressure at a given temperature and decreases with temperature at a given pressure, which is consistent with the trend of volume.

Fig.5(b) represent the temperature variation of Debye temperature $\Theta_D$ at zero pressure of $TaB_2$ and $AlB_2$, the inset of which shows the Debye temperature at 300K as a function of pressure. We observe that $\Theta_D$ decreases non-linearly with increasing temperature for both compound. On the other hand the pressure dependent Debye temperature exhibits a non-linear increase. We know that $\Theta_D$ is related to the maximum thermal vibration frequency of a solid. The variation of $\Theta_D$ with pressure and temperature assures that the thermal vibration frequency of atoms changes with pressure and temperature.

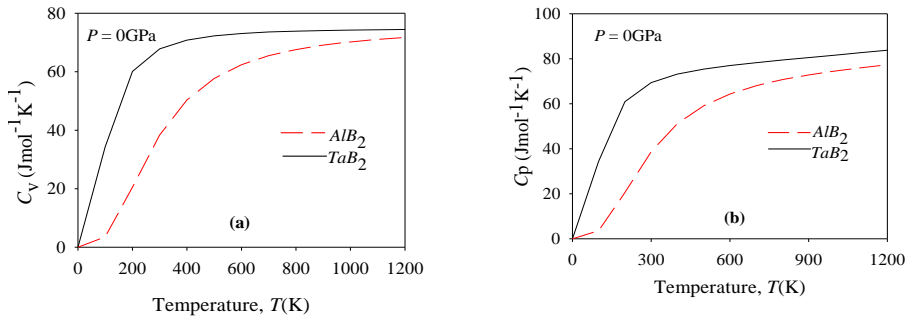

**Fig-6:** Temperature dependence of (a) specific heat at constant volume, and (b) specific heat at constant pressure of $TaB_2$ and $AlB_2$.

Fig.6 shows the temperature dependence of constant-volume and constant-pressure specific heat capacities $C_V$, $C_P$ of $TaB_2$ and $AlB_2$. The heat capacities increase with increasing temperature, because phonon thermal softening occurs when the temperature increases. The difference between $C_p$ and $C_v$ for $TaB_2$ and $AlB_2$ is given by $C_P - C_V = \alpha_v^2(T)BTV$, which is due to the thermal expansion caused by anharmonicity effects. In the low temperature limit, the specific heat exhibits the Debye $T^3$ power-law behavior and at high temperature ($T>550$ K for $TaB_2$ and $T>1000$ K) the anharmonic effect on heat capacity is suppressed, and $C_V$ approaches the classical asymptotic limit of $C_V=3nNK_B=$ 74 J/mol.K. This results show that the interactions between ions in the nanolaminates have great effect on heat capacities especially at low temperature.

Fig .7 shows the volume thermal expansion coefficient, α$_v$ as a function of temperature and pressure (inset). At low temperatures α$_v$ increases exponentially with temperature upto 300*K* for TaB$_2$ and 500*K* for AlB$_2,$ when the pressure is zero and gradually turns to a linear increase at high temperature. On the other hand, at a constant temperature α$_v$ decreases with increasing temperature. It is established that the volume thermal expansion co-efficient is inversely related to the bulk modulus of a material

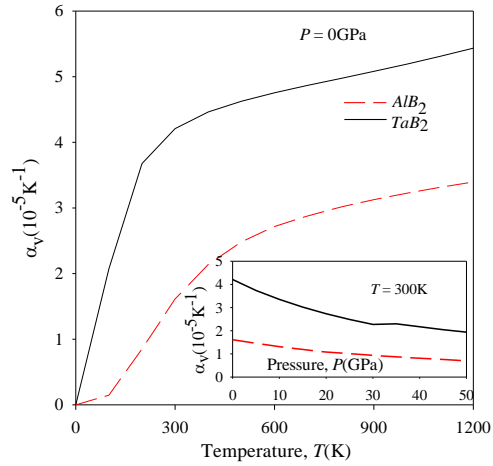

**Fig-7:** Temperature dependence of volume thermal expansion coefficients of TaB$_2$ and AlB$_2$. *Inset* shows pressure variation.

## 4. CONCLUSIONS

The structural, elastic, electronic, optical and thermodynamic properties of AlB$_2$ and TaB$_2$ have been investigated by using density functional theory. Elastic constants calculations reveal that the investigated structures are mechanically stable and they can be considered as stiff, anisotropic as well as brittle materials. Band structure results show that the compounds are metallic in nature. Absorption and conductivity spectrum shows metallic nature also. Furthermore, the dielectric function, energy-loss spectrum, absorption spectrum, conductivity and reflectivity were obtained and discussed in detail. The large reflectivity of TaB$_2$ in the low-energy region indicates suitability of the compound for use in practical purposes to remove solar heating. The temperature and pressure dependence of bulk modulus, specific heats, Debye temperature and thermal expansion coefficient are investigated by the quasi-harmonic Debye model and the results are discussed. The variation of Θ$_D$ with temperature and pressure reveals the changeable vibration frequency of the particle in AlB$_2$ and TaB$_2.$ The increase of heat capacity with increasing temperature shows that phonon thermal softening occurs when the temperature increases. Finally, our results are compared with the experimental and theoretical data where available. We hope that our calculated results could serve as a reference for future experimental study on the optical properties of the both phases.

**ACKNOWLEDGEMENT**

The authors would like to express their thanks to Professor A. K. M. Azharul Islam, Department of Physics, Rajshahi University, Rajshahi-6205, Bagnladesh for computational support at his laboratory.